\documentclass[aps,prl,preprint,groupedaddress,showpacs]{revtex4}
\usepackage{graphicx}
\begin{document}

\title{Reply  to " Comment on 'Once more about the $K\bar{K}$
molecule approach to the light scalars' " }
\author {
N.N. Achasov} \email{achasov@math.nsc.ru}
\author {
A.V. Kiselev} \email{kiselev@math.nsc.ru}

\affiliation{
   Laboratory of Theoretical Physics,
 Sobolev Institute for Mathematics, Novosibirsk, 630090, Russia}

\date{\today}

\begin{abstract}
The need to regularize loop integrals in a manner that preserves
     gauge invariance, for example, using the Pauli-Villars method, requires
     a subtraction that in the large mass limit hides its high momentum
     origin. This gives rise to the illusion  that only nonrelativistic
     kaon loop momenta are relevant, when in fact this is not the case,
     as we show.
 \end{abstract}


\maketitle

The authors of Ref. \cite{Comment} adduced the following argument
against our criticism \cite{achasov2007} . Since   the $\phi\to
K^+K^-\to\gamma (a_0/f_0)$ amplitude vanishes for gauge invariance
when the photon momentum vanishes, only those terms of the
integrand have the physical sense which vanish with vanishing  the
photon momentum. What actually happens is that cancellation of
contributions from different places of momentum (or co-ordinate)
space is realized. It is commonplace in electrodynamics. In
particular, low energy theorems are based on this. Discarding the
integrand in the third term of Eq. (3), the second term in Eq.
(4), and the contribution of Eq. (5) the authors of \cite{Comment}
distort the physical significance of the $K^+K^-$ loop model
because these contributions represent the high momentum and charge
flow distributions of kaons.
 Below we show
that the $K^+K^-$ loop model describes the relativistic physics.

\begin{figure}[h]
\begin{center}
\begin{tabular}{ccc}
\includegraphics[width=3.5cm]{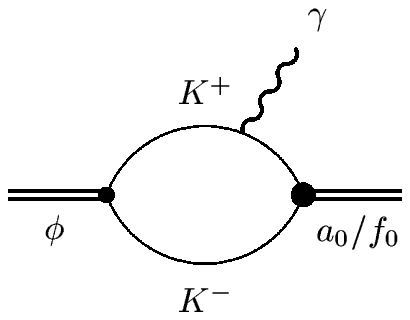}& \raisebox{-6mm}{$\includegraphics[width=3.5cm]{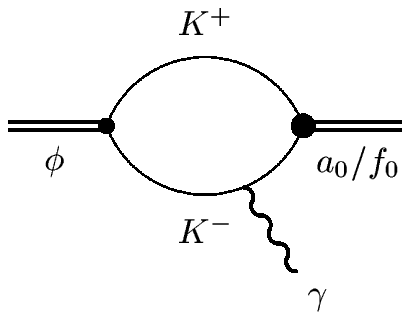}$}&
\includegraphics[width=3.5cm]{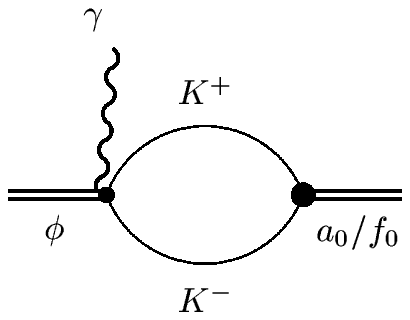}\\ (a)&(b)&(c)
\end{tabular}
\end{center}
\caption{Diagrams contributing to the radiative decay
amplitude.\label{abc}}
\end{figure}

When basing the experimental investigations of the light scalar
mesons production in the $\phi$ radiative decays $\phi\to \gamma
\bigl [a_0(980)/f_0(980)\bigr]\to\gamma \bigl[(\pi^0\eta
)/(\pi^0\pi^0)\bigr]$, there was  suggested \cite{achasov1989} the
kaon loop model $\phi\to K^+K^-\to\gamma \bigl [
a_0(980)/f_0(980)\bigr ] $ with the pointlike interaction, see
Fig. \ref{abc}.
 This model is used in
the data treatment and  ratified by experiment. In Refs.
\cite{achasov2001,achasov2003} an analysis of mechanisms of decays
under consideration was  carried out, which gave the clear
arguments for this kaon loop model.

Every diagram contribution in
\begin{equation}
 T\left \{\phi(p)\to\gamma [a_0(q)/f_0(q)]\right \}=
 (a)+(b)+(c)
 \label{T}
\end{equation}
is divergent hence should be regularized in a gauge invariant
manner, for example, in the Pauli-Wilars one.

\begin{eqnarray}
&&\overline{T}\left \{\phi(p)\to\gamma [a_0(q)/f_0(q)], M\right\}
= \overline{(a)}+\overline{(b)}+\overline{(c)}\,,\nonumber\\
&&\overline{T}\left\{ \phi(p)\to\gamma [a_0(q)/f_0(q)], M
\right\}=\epsilon^\nu(\phi)\epsilon^\mu(\gamma)\overline{T}_{\nu\mu}(p,q)\nonumber\\
&&=\epsilon^\nu(\phi)\epsilon^\mu(\gamma)\left[\overline{a}_{\nu\mu}(p,q)+\overline{b}_{\nu\mu}(p,q)+
\overline{c}_{\nu\mu}(p,q)\right],
 \label{overT}
\end{eqnarray}
\begin{eqnarray}
&&\overline{a}_{\nu\mu}(p,q)= -\frac{i}{\pi^2}\int\Biggl
\{\frac{(p-2r)_\nu(p+q-2r)_\mu}{(m_K^2-r^2)[m_K^2-(p-r)^2][m_K^2-(q-r)^2]}\nonumber\\[6pt]
&&-\frac{(p-2r)_\nu(p+q-2r)_\mu}{(M^2-r^2)[M^2-(p-r)^2][M^2-(q-r)^2]}\Biggr
\}dr\,, \label{overanm}
\end{eqnarray}
\begin{eqnarray}
&&\overline{b}_{\nu\mu}(p,q)= -\frac{i}{\pi^2}\int\Biggl
\{\frac{(p-2r)_\nu(p-q-2r)_\mu}{(m_K^2-r^2)[m_K^2-(p-r)^2][m_K^2-(p-q-r)^2]}\nonumber\\[6pt]
&&\hspace*{-6pt}-\frac{(p-2r)_\nu(p-q-2r)_\mu}{(M^2-r^2)[M^2-(p-r)^2][M^2-(p-q-r)^2]}\Biggr
\}dr\,, \label{overbnm}
\end{eqnarray}
\begin{equation}
\overline{c}_{\nu\mu}(p,q)=-\frac{i}{\pi^2}\, 2g_{\nu\mu}\int dr
\Biggl \{\frac{1}{(m_K^2-r^2)[m_K^2-(q-r)^2]}-
\frac{1}{(M^2-r^2)[M^2-(q-r)^2]}\Biggr \}\,,
 \label{overcnm}
\end{equation}
where $M$ is the regulator field mass. $M\to\infty$ in the end
\begin{equation}
\overline{T}\bigl [\phi\to\gamma (a_0/f_0), M\to\infty\bigr ]\to
T^{Phys}\bigl [\phi\to\gamma (a_0/f_0) \bigr ]\,. \label{Tphys}
\end{equation}
 We can shift  the
integration variables in the regularized amplitudes and easily
check the gauge invariance condition
\begin{equation}
\epsilon^\nu(\phi)k^\mu\overline{T}_{\nu\mu}(p,q)=\epsilon^\nu(\phi)(p-q)^\mu\overline{T}_{\nu\mu}(p,q)=0\,.
\label{gauge}
\end{equation}
 It is instructive to consider how the gauge
invariance condition
\begin{equation}
\epsilon^\nu(\phi)\epsilon^\mu(\gamma)\overline{T}_{\nu\mu}(p,p)=0
\label{pp}
\end{equation}
holds true,
\begin{eqnarray}
&&\epsilon^\nu(\phi)\epsilon^\mu(\gamma)\overline{T}_{\nu\mu}(p,p)=\nonumber\\
&&\epsilon^\nu(\phi)\epsilon^\mu(\gamma)T^{\,m_{K}}_{\nu\mu}(p,p)
-\, \epsilon^\nu(\phi)\epsilon^\mu(\gamma)T^M_{\nu\mu}(p,p)=
(\epsilon(\phi)\epsilon(\gamma))(1-1)=0\,.
\label{0pp}
\end{eqnarray}

 The
superscript $m_K$  refers to the nonregularized amplitude and the
superscript $M$
 refers to the regulator field amplitude.
 So, the contribution of the (a), (b), and (c) diagrams does not
depend on a  particle mass in the loops ($m_K$ or $M$) at $p=q$
\cite{typical}. But, the physical meaning of  these contributions
is radically different. The
$\epsilon^\nu(\phi)\epsilon^\mu(\gamma)T^{\,m_{K}}_{\nu\mu}(p,p)$
contribution is caused by  intermediate momenta (a few GeV)  in
the loops , whereas the regulator field contribution is caused
fully by high momenta ($M\to\infty $) and teaches us how to allow
for high $K$ virtualities in a gauge invariant way.

Needless to say the integrand of
$\epsilon^\nu(\phi)\epsilon^\mu(\gamma)\overline{T}_{\nu\mu}(p,p)$
is not equal to 0.

 It is clear that
\begin{eqnarray}
&&\epsilon^\nu(\phi)\epsilon^\mu(\gamma)T^{M\to\infty}_{\nu\mu}(p,q)\to
\epsilon^\nu(\phi)\epsilon^\mu(\gamma)T^{M\to\infty}_{\nu\mu}(p,p)\nonumber\\
&&\equiv\epsilon^\nu(\phi)\epsilon^\mu(\gamma)T^{M}_{\nu\mu}(p,p)\equiv(\epsilon(\phi)\epsilon(\gamma))\,.
\label{sc}
\end{eqnarray}

  So, the  regulator field contribution
tends to the subtraction constant when $M\to\infty $.

The finiteness of the subtraction constant hides its high momentum
origin  and gives rise to an illusion of a nonrelativistic physics
in the $K^+K^-$ model with the pointlike interaction. See, for
example,  Ref. \cite{kalash}; see Section 2 in this paper.

 This work was supported in part by  the
Presidential Grant No. NSh-1027.2008.2 for  Leading Scientific
Schools and by  RFFI Grant No. 07-02-00093 from the Russian
Foundation for Basic Research. A.V.K. thanks very much the Dynasty
Foundation and ICFPM for support, too.


\begin{thebibliography}{99}
\bibitem{typical}
A typical example of such integrals is
$2\int_0^\infty\frac{m^2x}{(x+m^2)^3}dx=1$.
\end{thebibliography}
\end{document}